\documentstyle[12pt,epsf]{article}
\input epsf
\def\beq{\begin{eqnarray}}   \def\eeq{\end{eqnarray}}
\def\pf{M_{\rm Pf}}

\begin{document}
\begin{flushright}
NYU-TH/00/06/05 \\
June 1, 2000
\end{flushright}

\vspace{0.1in}
\begin{center}
\bigskip\bigskip
{\large \bf On Sub-Millimeter Forces From Extra Dimensions} 

\vspace{0.5in}      

{Gia Dvali, Gregory Gabadadze, Massimo Porrati}
\vspace{0.1in}

{\baselineskip=14pt \it 
Department of Physics, New York University,  
New York, NY 10003 } 
\vspace{0.2in}
\end{center}

\vspace{0.9cm}
\begin{center}
{\bf Abstract}
\end{center} 
\vspace{0.1in}

We show that in theories with large extra dimensions forces
mediated by a bulk dilaton and bulk gauge fields may be parametrically
(exponentially) weaker than gravity due to the 
suppression of their wave-functions on a  brane. 
This is the case when dilaton gets stabilized by  certain 
strongly coupled dynamics on the brane, or the bulk gauge symmetries are
spontaneously broken by the Higgs mechanism on the worldvolume.
At distances smaller than  the size of a largest extra dimension these 
particles produce the force-law which decreases with distance faster 
than high-dimensional gravity. 
For a millimeter size extra dimensions
predicted deviations are in the range which  
may be detected in sub-millimeter gravity measurements.

\newpage

\vspace{0.2in}
{\bf  1. Introduction}
\vspace{0.1in} \\

{~}Theories with low-scale quantum gravity 
provide an additional motivation to  study  the physics at sub-millimeter
distances.
In this  framework the ordinary matter is localized on a brane 
which is embedded
in space with $N$ large dimensions to which gravity can spread 
\cite {ADD,ADD1}. 
As a result, the four-dimensional gravity measured at large distances is 
weak. These theories predict a deviation from Newtonian potential
at distances
below the maximal size of extra dimensions, $R~ <~{\rm mm}$. 
The deviation can be measured in principle by  
upcoming experiments, 
provided that $R$ is somewhere in the range of a tenth of a micron or so
\cite {Price,Kapitulnik}.

In the present paper we study the 
forces which, in addition to gravity,  are  present 
in the class of theories mentioned above. 
These forces  can be tested and constrained 
by ``table-top'' experiments even for
$R << $ mm \cite {Price,Kapitulnik}. 
Examples of these are the forces  mediated by some of the 
bulk fields. The bulk fields, by definition, 
can propagate in extra space and can couple to
matter on the brane with the strength which is 
comparable to that of  gravity. 
This statement is independent of the number and size of extra
dimensions. 
The most widely discussed  extra forces are 
those mediated by a dilaton and by bulk gauge fields. 
In the massless limit  the coupling of their zero-modes
to matter fields on a brane is suppressed by an universal
volume factor
\begin{equation}
{1 \over \sqrt{M_{\rm Pf}^{2 + N}V_N}}~. 
\label{coupling}
\end{equation}
Here, $M_{\rm Pf}$ denotes the fundamental Planck constant of 
the $(4+N)$-dimensional theory, and $V_N$ stands for 
the volume of extra compactified space.
It is assumed usually  that these particles interact with 
matter fields with  the  
coupling which is comparable to (in the case of a dilaton) or even 
stronger than (in case of gauge interactions) \cite{ADD1}
the gravitational one. If the physics responsible for their masses
originates from the brane,  these particles  acquire 
masses which are typically 
of order an inverse  millimeter or less.
Therefore,  it is expected that at sub-millimeter distances they
can mediated forces comparable or much stronger than gravity \cite{ADD1}.

In the present note we argue that this lore is
violated in certain cases:

\begin{itemize}

\item{Interactions of bulk states with ordinary matter fields can
be parametrically 
(exponentially) suppressed, so that forces mediated by these states can be
considerably weaker than gravity;}

\item {At distances less than the size of the extra dimension
they can give rise to  a force-law which  differs  
from that of gravity}. 

\end{itemize}

The origin of these phenomena is quite simple. 
If the masses of the bulk  states
are induced due to interactions on a  brane, then the wave-functions
of the corresponding lowest modes are strongly suppressed in the  
brane worldvolume.
In other words, as in ordinary quantum-mechanical systems,
the corresponding  wave-function ``sees'' the  mass term induced on a brane as
a repulsive potential and tends  to avoid it.
This is true for  both a dilaton and bulk gauge fields.
Concrete  examples will be given in the subsequent sections. 
In the case of bulk gauge fields there is a simple 
physical interpretation of the effect. Let us discuss this
interpretation first.  

Consider $U_B(1)$ gauge symmetry in the bulk. We will be assuming that
(some of the) standard model matter fermions are charged under this
group. In particular, this may be a gauged $(B -L)$-symmetry or some other
combination of lepton and baryon numbers.  The
corresponding ``photon'' $A_{B}$ can propagate in the bulk and can interact
with the matter localized on the brane. If a  zero-mode of the gauge
field is massless, then there is an unbroken gauge symmetry in the 
effective 4D worldvolume theory with an effective gauge 
coupling constant given by \cite{ADD1}
\begin{equation}
 g_{\rm eff} = {1 \over (\pf R)^{{N \over 2}}} = {\pf \over M_P}~,
\label{gc}
\end{equation}
were we assume that a high-dimensional gauge coupling is of the order of
unity  in the $\pf$-units\footnote{This is realized if, for instance, 
a  3-brane is embedded in a worldvolume of a higher 
dimensional brane on which the dimensionfull gauge coupling is defined by the 
string scale.}.
This is the strength by which an each Kaluza-Klein (KK)  
mode of the bulk gauge field $A_{B}$
couples to the $U_B(1)$-charged matter, in particular, the matter 
on a brane.
The force mediated by these interactions between, say, two protons
is related to their gravitational attraction as follows: 
\begin{equation}
 {F_{\rm gauge} \over F_{\rm gravity}}~\propto ~ {\pf ^2 \over m_{p}^2}~,
\label{gaugegravi}
\end{equation}
were $m_{p}$ is the proton mass.  For $\pf  \sim (1 -10)$ TeV, this 
gives the force  which is  
$(10^{6}-10^{8})$ times bigger than that of gravity.
Let us now suppose  that the $U_B(1)$
symmetry is spontaneously broken by some Higgs field $\chi$ 
(not to be confused with the Standard Model Higgs)
which condenses on a brane. 
A simple argument based on 4D gauge invariance
suggest that the gauge field acquires  the  mass
\begin{equation}
m_{\rm gauge}~ =~ g_{\rm eff}\langle \chi \rangle~,
\end{equation}
and the gauge force will be screened at distances $r >> m^{-1}_{\rm gauge}$.
On the other hand, at distances $r << m^{-1}_{\rm gauge}$ 
charges are  not screened
and power-law gauge forces can be observed.  One may expect, therefore,
that the force at smaller distances will again be related to 
the gravitational force as in (\ref{gaugegravi}). 
However, as we will see,  this expectation is violated
if $\chi$ lives on ``our'' brane. In fact, 
even at distances $r << m^{-1}_{\rm gauge}$, the force may be
{\it exponentially} weaker than gravity.
The effect can be most easily understood by means of the analogy with 
superconductivity.
As seen by  a  high-dimensional observer, the brane becomes superconducting,
and the photon is repelled from 
its worldvolume due to the  high-dimensional version
of the Meissner effect. As a result, the coupling of the
photon to fields localized inside the superconductor is suppressed
by the strength of the condensate. This is an additional suppression,
which in fact is  absent for strictly massless 
high-dimensional fields, in particular for gravity. 
Therefore, the interaction mediated by a bulk gauge field on the brane
is expected to  be parametrically suppressed with respect to gravity.
In the next two sections we will demonstrate this on concrete examples.

\vspace{0.2in}
{\bf 2. Extra dimensions in the infinite volume limit}
\vspace{0.1in} \\

Let us first consider the dilaton field. 
It enters the action in higher dimensional space  as follows:
\beq
\pf^{2+N}~\int d^{4+N}X\sqrt{|G|}\left ({\cal R}~+~G^{AB}
\partial_A \Phi\partial_B \Phi~+...\right )~.
\label{Action}
\eeq
Here, ${\cal R}$ is the $(4+N)$-dimensional Ricci scalar, and $...$ 
stand for other possible terms and interactions.
The dilaton field $\Phi$ interacts with the states 
which live  in the brane worldvolume. This interaction in general 
takes the form:
\beq
S_{\rm int}~=~\int d^4x~\sqrt{|G(x,0)|}~\Phi(x,0)~{\cal O}(x)~,
\label{int}
\eeq
where $x$ denotes the 4D coordinates, $G(x,0)$ is the induced metric
on a brane, and  
${\cal O}(x)$ denotes  an operator of the brane worldvolume 
theory. An example of this operator would be the trace of the energy-momentum 
tensor of the worldvolume theory, $T^\mu_\mu(x)$.
If the dilaton were massless, its exchange would produce scalar
attraction of the gravitational strength, 
which would be phenomenologically  unacceptable.
Therefore, a dilaton should be massive in any physically viable 
scenario. Usually it is assumed that the dilaton mass comes from some
non-perturbative dynamics which generates its potential. Let us ask
what happens if this potential is generated due to the physics localized
on our brane. The precise nature of this mass will be irrelevant for 
our analysis.
We would like to investigate the potential mediated by the dilaton 
on a brane at the distances smaller
than the compactification  radius. Therefore, as a first step we will  
study  one uncompactified extra dimension.

Suppose  that the  dilaton potential $V(\Phi)$ 
is generated on a brane which is located in extra space
at the point $y=0$. Then the action for the dilaton 
can be written as follows:
\beq
{\pf^{3}\over 2}~\int d^{4}x~dy \left \{  
\partial_A \Phi\partial^A \Phi~-~V(\Phi)~\delta(y)~+...~\right \}~.
\label{V}
\eeq
In this expression, for sake of simplicity, we omit the 
gravitational part; it is assumed that gravity 
produces an appropriate classical background on which the  
dilaton field propagates (the dilaton background will be discussed below). 
The corresponding classical equation of motion
for the dilaton takes the form:
\beq
~\partial_B \partial^B ~\Phi~+ ~\delta(y)~
{dV\over d\Phi}~ = ~0~.
\label{eom}
\eeq
The solution to this equation  defines the classical 
dilaton background which we denote by $\Phi_{\rm cl}$.
We are interested in fluctuations about this background.
Therefore, we decompose  the dilaton field as follows:
\beq
\Phi~=~\Phi_{\rm cl}~+~{\Delta  \Phi}~.
\label{fluctuations}
\eeq 
The equation of motion for the fluctuations takes the form:
\beq
\left (~\partial_B \partial^B ~+ ~\delta(y)~
{d^2V\over d\Phi^2}|_{\Phi_{\rm cl}}~\right)~{\Delta \Phi}~ = ~0~.
\label{eomP}
\eeq
The quantity ${d^2V/ d\Phi^2}|_{\Phi_{\rm cl}}$ can be thought 
of as the dilaton ``mass'' term on the brane\footnote{Note 
that we use an unconventional normalization  
for the field $\Phi$.}. In what follows we denote this 
quantity by $M\equiv {d^2V/ d\Phi^2}|_{\Phi_{\rm cl}}$.
Thus, 
the dilaton fluctuations obey the following
linearized equation: 
\begin{equation}
\left(~\partial_B^2 ~+ ~M\delta(y)~\right)~{\Delta \Phi}(x,y)~ = ~0~.
\label{dilaton}
\end{equation}
This reduces to a simple quantum-mechanical problem.
Indeed, introducing the notations:
\beq
~{\Delta \Phi}(x,y)~=~e^{ipx}~\Psi_m(y)~, ~~~~~p^2~\equiv~p_0^2~-~
{\vec p}^{~2}~=~m^2~,
\eeq
we derive the following Schr\"odinger type equation:
\beq
\left(-~\partial_y^2 ~+ ~M\delta(y)~ \right)~\Psi_m(y)~ = ~m^2~\Psi_m (y)~.
\label{schrod}
\eeq
This equation has a continuous  tower of 
solutions with plane wave profiles in the extra dimension.
The quantity which we will need 
below is the square of the wave-function on the brane. This looks as
follows:
\begin{equation}
\left | \Psi_m(y=0) \right |^2~= ~{4~m^2 \over 4~m^2 ~+~ M^2}~.
\end{equation}
We see that the wave-functions of the light modes 
(which are most important at
large distances) are suppressed by the factor $m^2/M^2$. The
exchange of each mode mediates the  Yukawa type force
\begin{equation}
{{\rm exp}({-m~r})\over r}~,
\end{equation}
among the particles localized on the brane (here $r$ denotes a 3D distance
on a brane). The net force is obtained
by integrating over  $m$. However, the leading  effect comes 
only from  the modes which are lighter  than $M$. The wave functions of other
states are unsuppressed on the brane and they produce  the 
usual higher-dimensional
potential. 
The leading term in the potential mediated between two 
non-relativistic particles of masses 
$m_1$ and $m_2$ is equal to
\beq
V(r)~ \simeq  ~m_1~m_2 ~G_5 \int_0^M  
~{{\rm e}^{-mr}\over r}\left |\Psi_m(y=0) \right |^2 dm
\nonumber \\ \nonumber \\
~\simeq~{4 ~m_1~m_2 ~G_5 \over M^2} 
\int_0^M  {{\rm e}^{-mr}\over r}~ m^2~dm~
\propto ~{m_1~m_2~ G_5 \over M^2~r^4}~+...~~
\eeq
Here, $G_5$ denotes the 5D Newton constant and dots stand for 
subleading corrections.
We see a very strong additional suppression (compared to the 
5D gravitational law) by a factor $M^2r^2 >> 1$. 
This indicates that dilaton-mediated force is much weaker than 
gravity.

Given the significance  of this result it is instructive to understand it 
in a different approach. We can  calculate  
the corresponding retarded Green  function for the dilaton 
and then find  the potential. 
The classical equation for the Green  function  looks as follows:
\beq
\left ( \partial_A\partial^A~+~M~\delta(y)~  
\right ) ~G_R(x,
y; 0,0)~=~\delta^{(4)}(x) \delta(y)~,
\label{green}
\eeq
where $G_R(x,y; 0,0)=0$ for $x_0<0$.
The potential mediated by the dilaton 
on the 4D worldvolume of the brane is determined as:
\beq
V(r)~=~\int G_R\left (t,\vec{x},y=0; 0,0,0\right ) dt~,
\label{pot}
\eeq
where $r\equiv\sqrt{x_1^2+x_2^2+x_3^2}$. 
To find a solution of (\ref {green}) let us turn to 
Fourier-transformed quantities with respect to 
the worldvolume four-coordinates $x^\mu$:
\beq
G_R(x,y; 0,0)~\equiv~\int ~ {d^4p\over (2\pi)^4}~e^{ipx} 
~{\tilde G}_R(p,y)~. 
\label{Fourie}
\eeq 
Turning to Euclidean space the equation (\ref {green})
takes the form:
\beq
\left (~(p^2-\partial_y^2)~+~M~\delta(y) ~\right )~  
{\tilde G}_R(p,y)~=
~\delta(y)~. 
\label{mom}
\eeq
Here $p^2$ denotes the square of an Euclidean four-momentum.
The solution with boundary conditions appropriate for our problem 
takes the form:
\beq
{\tilde G}_R(p,y)~=~{1\over M~+~2 p}~ {\rm exp} (-p|y|)~,
\label{sol1}
\eeq
where $p\equiv\sqrt {p^2}=\sqrt{p_4^2+p_1^2+p_2^2+p_3^2}$. 
Using this expression and Eq. (\ref {pot}) one finds the following
formula for the  4D dilaton-mediated potential:
\beq
V(r)\propto~
{1\over 2r}~\left ( -{d\over dr} \right )~{\rm Re}~\left \{ {\rm exp}
\left (-{iMr\over
2} \right )
E_1\left (-i{Mr\over 2}\right )  \right \}~.
\eeq
Here, $E_1(z)\equiv \int_z^\infty e^{-t}dt/t$ denotes the exponential integral.
For the large distances this gives
\beq
V(r)~\propto~{1\over M^2~r^4}~.
\eeq
Therefore, we find the same $1/r^4$ scaling of the potential
at large distances.  We conclude that the dilaton-mediated
potential differs  from the gravitational one
due to the suppression of the dilaton wave-function on a brane.

So far we have studied  the case when the  dilaton 
potential $V(\Phi)$ is generated
on a brane.   It is 
conceivable that a kinetic term for a dilaton can also be induced  
on a brane (along the lines of Ref. \cite {DGP3}). In this case,
the scaling laws of the force mediated by a dilaton would  
change. Indeed, let us assume that  there is an additional  
kinetic term in (\ref {V}) of the following form:
\beq
\pf^3~{\partial_\mu~\Phi~\partial^\mu~\Phi \over 2~\Lambda}~\delta(y)~,
\eeq 
where $\Lambda$ is some dimensionfull parameter. 
With this term in (\ref {V}) we obtain the following expression for
the Green function (\ref {sol1})
\beq
{\tilde G}_R(p,y)~\propto~
{\Lambda \over p^2~+~2\Lambda ~p~+~M~\Lambda}~ {\rm exp} (-p|y|)~.
\label{sol2}
\eeq
At short distances, $r<<{\rm min}\{1/\sqrt{M\Lambda},~1/\Lambda \}$, 
the corresponding potential scales as $1/r$. 
In the region 
${\rm max}\{ 1/\sqrt{M\Lambda},~1/\Lambda\} << r << ~1/M$, 
the potential behaves as $1/r^2$.
Finally, at large distances
$r>>{\rm max}\{1/\sqrt{M\Lambda}, ~1/M \}$ the potential scales  as 
$1/r^4$; this term is further modified by subleading corrections.

\vspace{0.2in}
{\bf 3. Finite volume extra dimensions}
\vspace{0.1in} \\

In the previous section  we have treated the brane as a zero-thickness plane
embedded in the infinite-volume extra dimension.  Let us now study the case 
of  a finite-thickness brane embedded in a finite volume extra
space.  We shell denote  the brane thickness by  $B$ and will
assume that  the brane  is located
at the center of an interval. In this framework  
the details of the interactions
depend on the overlap between the transverse size  on which 
the dilaton potential is supported on a brane, 
and the transverse widths of the domain on which  
matter fields are localized. Generically this 
overlap can be small. 
In what follows, we will assume for simplicity that 
ordinary matter is localized at a single point in the transverse space,
i.e., we simply assume the  delta-function type 
localization at  $y = 0$. 
As we mentioned  above, the width of the transverse region on which
the dilaton potential is supported will be some nonzero
constant  $B$ (one could think of a brane on which matter fields
are localized, to be embedded into a wider brane on which 
the dilaton potential is supported). 
Qualitatively, this construction  can be modeled  by the 
following action
\beq
{\pf^3\over 2}~\int d^4x~dy ~\left \{ (\partial_A \Phi)^2~-~M_0^2~
\delta_R(y,B) ~\Phi^2
\right \}~+~\int d^4x ~\Phi(x,0)~ T^\mu_\mu(x)~+...
\label{action}
\eeq
Here, as before,  we  omit gravitons and other terms concentrating on
the dilaton field  only. $\delta_R(y,B)$
is a bell-shaped function which determines that the mass term
of the $\Phi$ field (or its potential) is spread over the transverse 
size $B$ of a  ``fat'' brane \footnote{In this case 
we think of the brane as a sort of ``kink'' for certain field theory.}. 
One can think of 
$\delta_R(y,B)$ as the smoothed Dirac's delta function 
with the unit hight and width $B$.
Let us compactify the extra space on a circle of  radius R.
We assume that $R$ is much bigger than the width 
$B$. Following the usual routine of compactification
we decompose the field $\Phi$ in its 4D and extra dimensional parts:
\beq
\Phi(x,y) ~=~ \sum_n \varphi_{n}(x)~\Psi_{n}(y)~.
\label{kk}
\eeq
In what follows, we will retain  for simplicity only the zero-mode;
all other modes can be included analogously. Setting the normalization
as 
\beq
\int_0^{2\pi R}~\Psi_n^2 (y)~dy~=2\pi ~R~,
\label{norm}
\eeq
and rescaling the zero mode
\beq
\varphi(x) ~=~~{\sigma \over \sqrt {2\pi ~R ~M^3_{\rm Pf}} }~=~
{\sigma\over \sqrt{M^2_P}}~=~\sqrt{G_N}~\sigma~,
\label{sigma}
\eeq
we get the canonically normalized Lagrangian
for the massive scalar field $\sigma$
\beq
{1\over 2}~\left (\partial_\mu\sigma \right )^2~-~{1\over 2}~
m^2~\sigma^2~+~
\sqrt{\Psi ^2(0)~ G_N}~\sigma~T^{\mu}_\mu~.
\label{siglag}
\eeq
Here the dilaton (would be) zero-mode 
mass is $m^2~=~M^2~\int~ \delta_R(y,B)~dy/R$. 
The interaction of the $\sigma$ field with
the matter is determined  not by the Newton constant alone, but also by 
$\Psi (0)$. Thus, the effective interaction constant reads 
as follows:
\beq
G_{\rm Dilaton}~=~G_N~\Psi ^2(0)~.
\eeq 
We will show below that $\Psi (0)$ is exponentially suppressed
\beq
\Psi ^2 (0)~\propto~{\rm exp} \left (-M_0B \right )~.
\label{chi}
\eeq
If the hight of the barrier is bigger than inverse
transverse size of the brane, then the suppression
coefficient is substantial. 
Therefore,  we discover that the coupling of the lowest mode to an ordinary
matter is exponentially suppressed, whereas it's mass is only power
sensitive to the width of the brane
\beq
 m_0^2~ \propto  ~{M_0^2~B\over R}~.
\label{mass}
\eeq 
Therefore, this force
can be exponentially weaker than gravity.
This effect has important
phenomenological implications.

Let us discuss the dilaton-mediated forces at scales $r << R$.
The interaction is mediated by the whole tower of the 
corresponding KK states. 
We should  compare this interaction with  the gravitational 
potential  at the same  distances.
The question  is whether the resulting potential is 
still suppressed after the summation of all the KK 
modes are performed. To find this out consider a simple toy model
in which the brane in compact space is approximated by a 
periodic array of square potential barriers. This is a good model 
for a kink with nonzero thickness. 
%\begin{figure}
%\centerline{\epsfbox{potential.ps}}
%\epsfysize=6cm
%\caption{\small The periodic array of potential barriers which
%model a ``fat''  brane in the compact extra dimension.} 
%\label{fig1}
%\end{figure} 
In the approximation of a very high barrier the wave function takes
the form:
\beq
\left | \Psi_{m_n}(y=0) \right |^2~=~{\cal G}(m_n,~R)~
~{\rm exp} \left ( -M_0~B \right )~,
\eeq
where
\begin{equation}
m^2_n~ = ~{n^2\over R^2}~ +~ m^2_0~,~~~~~{\cal G}(m_n,~R)~
\propto~{ {m_n}~\sin^2(m_n~R) ~({R~+~B})\over
{R~m_n-2\sin(2m_n~R)}}~,
\end{equation}
and the zero-mode mass is defined as $m^2_{0} = M_0^2~B/R$.
This demonstrates that wave functions for all KK states 
are exponentially suppressed as ${\rm exp}(-M_0B)$.
Furthermore, the effective interaction constants of these states with
the localized matter are exponentially suppressed as well.
 
Finally, we need to  determine the power-low-force which is produced
due to the exchange of all these modes. The problem  
simplifies computationally  if one considers the approximation  
of a single potential barrier with the width $B$ in  
infinite extra space. In this case,    
the potential produced by the tower of the KK states
reads as follows: 
\begin{equation}
V(r)~ \sim ~{m_1~m_2 ~G_5 ~{\rm exp}(-M~B) \over r^4~M^2}~+....
\end{equation}
Therefore, the same exponential suppression of the 
force mediated by the dilaton.

Phenomenologically, the most interesting for the sub-millimeter
gravitational measurements is the case of two or more extra dimensions.
Here the same qualitative features hold  as far  as 
the exponential sensitivity is concerned. 
Generically, the mass of the dilaton is
\begin{equation}
  m^2_{0}~\propto ~M_0^2 ~{B^N \over R^N}~,
\end{equation}
whereas the wave function of the light modes is suppressed
as $\sim {\rm exp}(-M_0B)$.

All the results obtained above for the dilaton  are equally applicable to
the bulk gauge fields  which acquire masses due to the Higgs effect on 
our brane (this Higgs effect should not be confused with 
the Standard Model Higgs effect.).
Therefore, the potential mediated by the exchange 
of the bulk gauge fields can exhibit the 
similar behavior:

(I) At distances $r>> \langle {\chi} \rangle ^{-1}$
($\chi$ - being the additional Higgs field giving masses 
to the bulk gauge bosons ) they can be parametrically 
(exponentially) suppressed compared to gravity; 

(II) At distances
$ \langle {\chi} \rangle^{-1} ~<<~ r ~<<~ R$ 
they can exhibit distinctive (from gravity)   
dependence on $r$ in the force law.

Finally, there might  be subtleties  related to the
couplings of  the longitudinal components of gauge fields.
Let us  address this issue.
Since we study the gauge fields which acquire masses via 
the Higgs effect on the ``fat'' brane, the would be Goldstone
boson (which  becomes a longitudinal component of the massive 
gauge field) is not a bulk mode. 
Therefore,  one might naively expect that the coupling of this mode
to the ordinary fermions is suppressed only by the relative thickness of 
the ``fat'' brane.  This is certainly  true if there
is a direct coupling between the standard model fermions and the 
Higgs field $\chi$ which  breaks a bulk gauge symmetry, and, moreover, 
its interactions, as, for instance, 
\beq
\chi~{\bar \Psi}~\Psi~,
\label{danger}
\eeq
are {\it not} invariant under independent $U_B(1)$ 
global transformations of $\chi$
\footnote{Note that if the bulk symmetry in question is used to
suppress dangerous operators (e.g.,  baryon number violating terms), 
such couplings are not desirable and must be suppressed by 
some discrete symmetries.
This is an additional motivation for not having them in the theory. 
However, for the sake of generality, we include these terms  into 
the  consideration.}.
If the terms similar to one in  
(\ref {danger}) are absent, there is no direct coupling of the
Goldstone mode to ordinary fermions. Indeed,   
the corresponding Higgs field could couple to fermions 
through invariant  couplings such as $\chi^*~\chi~{\bar \Psi}~\Psi~$.
In the gauge-less limit Goldstone
bosons would decouple. Thus, in this case,  the only possible 
coupling to the longitudinal
components of the gauge fields would be  through their transverse components,
which as we have demonstrated above,  are  exponentially suppressed.
On the other hand, the terms of the form
(\ref {danger}) 
would give rise to  a direct coupling of the standard model 
particles to a  low-scale Goldstone boson. 
There are severe astrophysical bounds on such couplings,
which would require a scale of the symmetry 
breaking to be at least $10^{10}$ GeV or so.
Therefore,  the phenomenological constraints 
force us to the scenario 
in which  the matter fermions have no direct couplings
to the Goldstone mode and, thus, to the longitudinal component
of the bulk gauge fields.

\vspace{0.2in}
{\bf 4. Conclusions}
\vspace{0.1in} \\

In conclusion, we have shown that in many cases bulk fields, such as
a dilaton or gauge fields, have parametrically  suppressed interactions with
ordinary matter on a brane. These fields  mediate forces which can   
be much  weaker than what is expected naively.
This fact should be taken into account while  deriving
possible experimental bounds on masses of the bulk
states  from the sub-millimeter gravitational 
experiments \cite{Price,Kapitulnik}.

\vspace{0.3cm}

{\bf Acknowledgments}

\vspace{0.2in}

We are grateful to Zurab Berezhiani and Marko Kolanovic
for useful discussions.
The work of G.D. is supported in part by a David and Lucille  
Packard Foundation Fellowship for Science and Engineering. G.G. is
supported by  NSF grant PHY-94-23002. 
M.P. is supported in part by NSF grant PHY-9722083.

\end{document}